\documentclass[prb,twocolumn]{revtex4-2}

\usepackage{graphicx}% Include figure files
\usepackage{xcolor}
\usepackage{dcolumn}% Align table columns on decimal point
\usepackage{bm}% bold math
\begin{document}

%\preprint{}

\title{Strain-induced superconducting proximity effect in topological insulator TaSe$_3$}

\author{R.M. Lukmanova}
\author{I.A. Cohn}
\author{V.E. Minakova}
 \author{S.V. Zaitsev-Zotov}
\email{serzz@cplire.ru}

\affiliation{Kotelnikov Institute of Radioengineering and Electronics of RAS, Mokhovaya 11, bld. 7, Moscow, 215009, Russia}

\affiliation{HSE University, Myasnitskaya Ulitsa, 20, Moscow, 101000, Russia}

\date{\today}

\begin{abstract}
The magnetoresistance of superconductor-topological insulator-superconductor structures, with indium as the superconductor and TaSe$_3$ as the topological insulator, shows steplike features on the resistance under magnetic fields. These resistance steps are resulted from the suppression of superconductivity, induced by the superconducting proximity effect in both the bulk and surface states of the topological insulator. The position and amplitude of the steps, occurring at approximately 0.1~T, show an unusual dependence on the magnitude of the uniaxial strain ($\epsilon$), indicating their connection with surface states. This behavior follows the expected transition sequence: semi-metal $\rightarrow$ strong topological insulator $\rightarrow$ trivial insulator, and supports the presence
of surface states at $0.46\% \lesssim \epsilon \lesssim 0.85\%$.
\end{abstract}

\maketitle
 
\section{Introduction}
The superconducting proximity effect is a phenomenon when a normal metal that is in close proximity to a superconductor acquires superconducting properties. 
The ability of the proximity effect to transfer superconducting properties into a normal metal or semiconductor opens up a wide range of potential applications in various fields of superconducting electronics and quantum technologies.

In s-wave superconductor - topological insulator (TI) structures, the superconducting proximity effect induces a state resembling a spinless superconductor and described in terms of Majorana zero modes \cite{fu_PRL2008_sc_ti,lutchyn2018majorana}. Majorana modes are non-local, interacting weakly with the environment, and are therefore considered as promising objects for use in qubits. For this reason, effects arising at the interface of a superconductor and a topological material (topological insulator, Weyl semimetal, etc.) have attracted increased attention in recent years \cite{sato2017topological,Proxti3,Proxti4,Proxti5,hart2014induced,pribiag2015edge,Esin2023gete,kononov2020one,ZhangPRB2011_Bi2Se3sc,williams2012unconventional}. 

When studying the superconducting proximity effect in topological materials, the structure under study is usually a strip of a two-dimensional TI \cite{hart2014induced,pribiag2015edge}, a Weyl semimetal \cite{Esin2023gete,kononov2020one}, or a whisker of a three-dimensional topological insulator \cite{ZhangPRB2011_Bi2Se3sc,williams2012unconventional}, in which superconductivity is induced due to the proximity effect using superconducting contacts. The close arrangement of the contacts leads to the flow of supercurrent, the formation of a superconducting circuit (2D TI) or a superconducting surface (3D TI), the properties of which are the object of study.

TaSe$_3$ is a quasi-one-dimensional transition metal trichalcogenide known since the mid-1960s \cite{otkr}. It has a chain-like structure with a monoclinic unit cell that belongs to a non-centrosymmetric orthorhombic space group P2$_1$/m \cite{otkr} and reveals metallic conductivity. 

Information about superconducting properties of TaSe$_3$ is controversial. On the one hand, some crystals of TaSe$_3$ go into a superconducting state at a temperature of about 2.2 K \cite{sambongi1977superconductivity,yamamoto_1978_sc,yamaya1981superconducting}, with filamentary superconductivity \cite{morita1987anisotropic}. On the other hand, no superconductivity was found in TaSe$_3$ crystals down to 50 mK \cite{monceau1977superconductivity}.

The electrical properties of this material turn out to be very sensitive to uniaxial strain which leads to a metal-insulator transition at a critical strain $\epsilon_{cr}\approx 1$\%\ \cite{tritt1986effect,lin_2021-strain_TI} (3\% according to \cite{hyun_2022_strain_TI}). Superconductivity 
in TaSe$_3$ is suppressed under uniaxial strain of 1\%\ \cite{tritt1986effect}.

Interest in this material has recently been reborn after the discovery of its topological properties \cite{Nie_TaSe3-topological_2018}. Density functional theory (DFT) calculations have shown that in the initial state, this material is a semi-metal due to the overlap of the valence and conduction bands, with a topologically protected state forming on its surface. As a result, it belongs to the class of topological insulators despite its metallic conductivity. According to the ARPES results, TaSe$_3$ is a weak topological insulator at deformations of $-2\% \leq \epsilon\leq -0.7$\%, a strong topological insulator in the range of $-0.7\lesssim \epsilon \leq 1$\%\, and passes into the topologically trivial state at $\epsilon\gtrsim 1$\% \cite{lin_2021-strain_TI}. According to the DFT results \cite{hyun_2022_strain_TI} TaSe$_3$ is initially in a weak topological phase, and go to a strong topological phase at $1\% \leq \epsilon\leq$ 3\%. To our knowledge, no other experimental evidence of the emergence and disappearance of surface states in TaSe$_3$ during deformation has been presented to date.

In this paper, we present the results of a study on the effect of deformation on the properties of In/TaSe$_3$/In structures. In these structures, resistance steps were observed in the resistance-magnetic field dependence at temperatures below the superconducting transition temperature of In. The appearance or disappearance of these steps depends on the deformation of the TaSe$_3$ crystal and exhibits an unusual dependence on the sample's resistance. The origin of these steps is attributed to the superconducting proximity effect on surface states, which develops only in a narrow region of strain.

\section{Methods}
High-purity TaSe$_3$ crystals were synthesized  by the chemical vapor transport method. A mixture of Ta and Se with a 10\%\ excess of Se relative to the stoichiometric amount was sealed into a quartz ampule and a temperature gradient of 670-730~$^o$C was applied for 10 days. The grown crystals have a needle-like shape, which is convenient for the study of strain effects on transport properties. No trace of superconductivity is observed in the synthesized crystals at temperatures above 1.1~K.

\begin{figure}
\includegraphics[width=\linewidth]{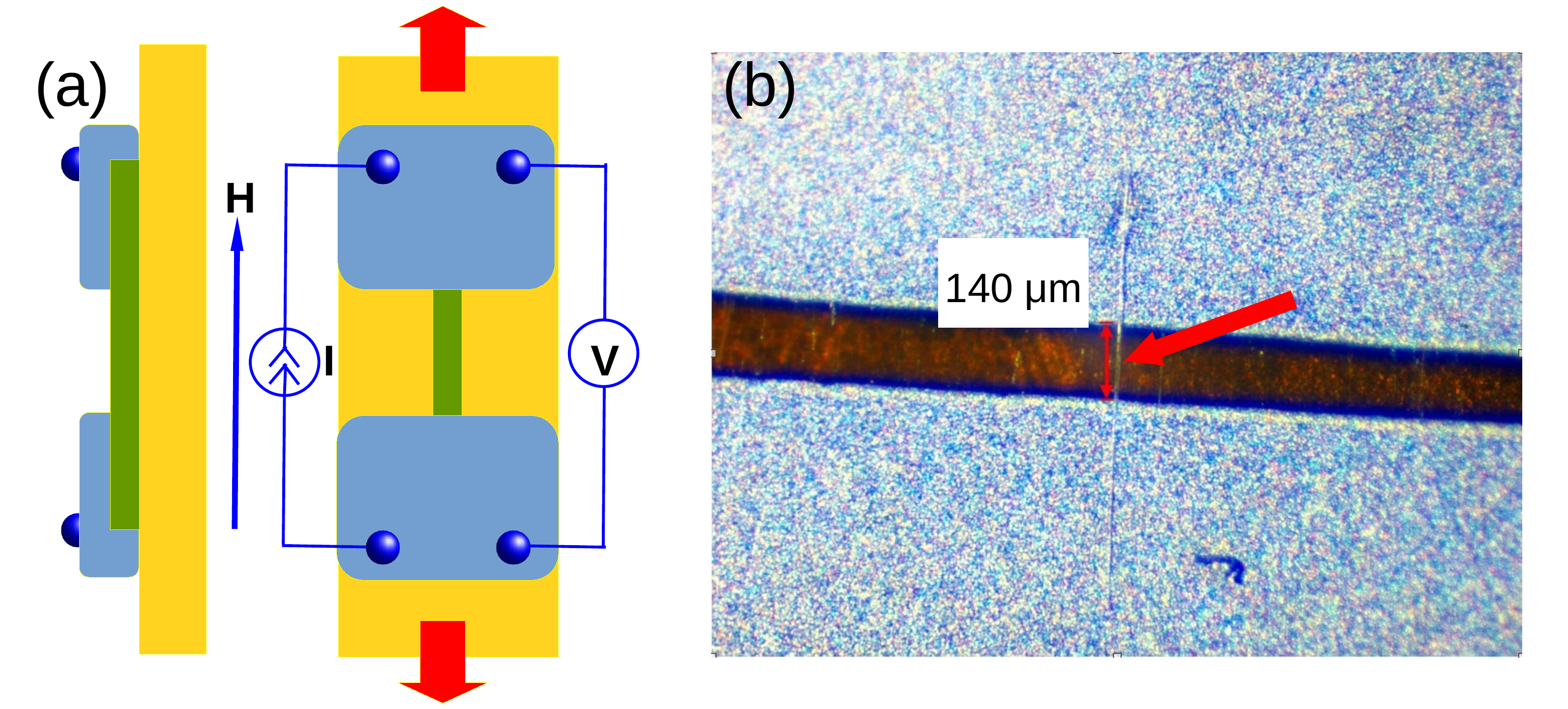}
\caption{(a) Schematic of the strain device. (b) Photograph of a 140 $\mu$m long sample with deposited indium contacts. Side illumination is used to make the crystal visible. The red arrow shows points to the sample.}
\label{fig:scheem}
\end{figure}

The device under study consists of a flexible substrate with a TaSe$_3$ crystal the ends of which are covered by two indium contacts (Fig.~\ref{fig:scheem}). The crystals were placed on a 100-$\mu$m-thick polyimide substrate. Indium contacts about 0.1-0.5 $\mu$m thick were deposited in a vacuum of $6\cdot 10^{-7}- 2\cdot 10^{-6}$~Torr through a mask. Typical dimensions of the samples: the distance between the contacts 50-150 $\mu$m, cross-sectional area of studied crystals are 0.05-0.5~$\mu$m$^3$. The strain was created by stretching the substrate (Fig.~\ref{fig:scheem}(a)) (see Ref.~\cite{mina_2023_strain} for further details).
The resistance $R$ of the obtained structures was measured by the four-contact method  as shown in Fig.~\ref{fig:scheem}(a). The data described below were obtained in a representative sample with the length $L=100$~$\mu$m, $R_{300K}(\epsilon =0)=10^4{\rm~}\Omega$ and cross-sectional area 0.05~$\mu$m$^2$.

\section{Results}
\begin{figure}
\begin{center}
\includegraphics[width=\linewidth]{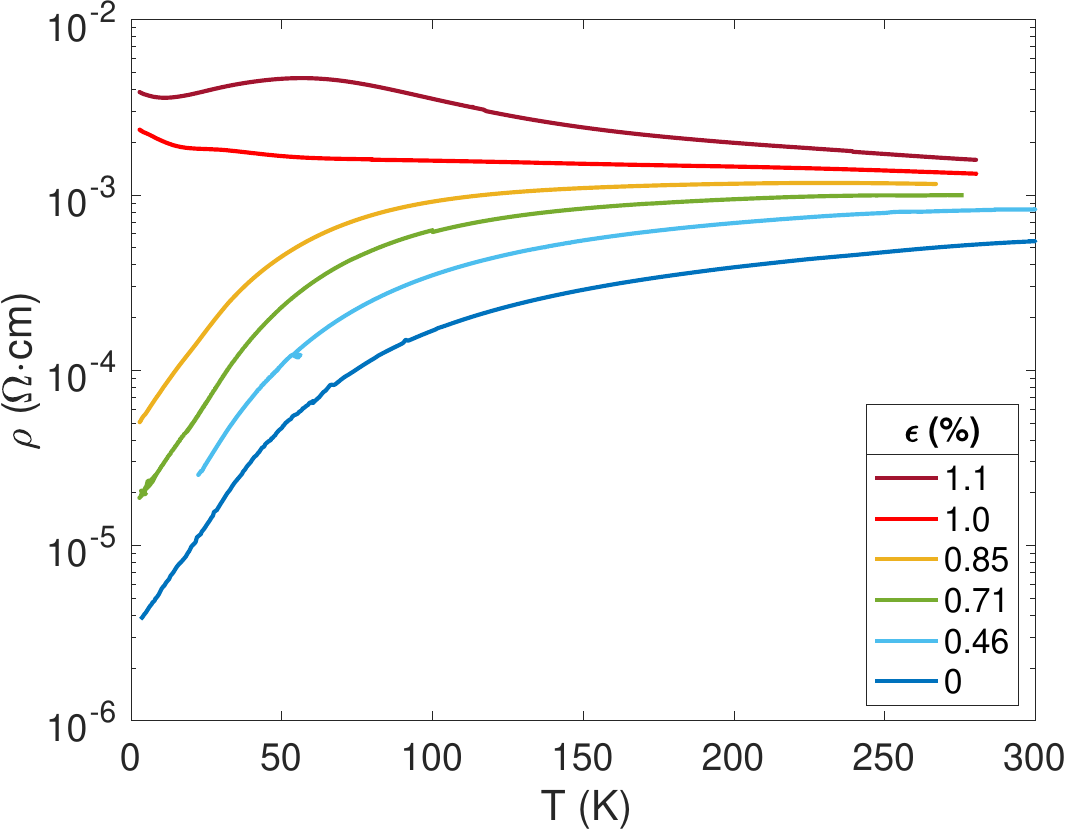}
\caption{Temperature dependencies of resistivity at different degrees of sample deformation at room temperature. Small steps on the lower curve are due to too rapid a change in temperature.
}
\label{fig:rt}
\end{center}
\end{figure}

Fig. \ref{fig:rt} shows temperature dependencies of the resistance of the representative TaSe$_3$ sample at different levels of uniaxial strain. In the initial state, the resistance of the sample is metallic with $RRR\equiv R_{300K}/R_{4.2K}>100$, which indicates the high quality of TaSe$_3$ crystals. With increasing strain, the resistance of the sample increases over the entire temperature range. As the strain approaches $\epsilon_{tr} \approx 1$\%, a transition to the insulator state occurs, accompanied by a significant increase in resistance at low temperature. In the region of strain near this transition, $\epsilon \gtrsim \epsilon_{tr}$, some $R(T)$ dependencies have a non-monotonic form with a maximum at a temperature of 70-120~K.

\begin{figure}
\begin{center}
\includegraphics[width=0.43\textwidth]{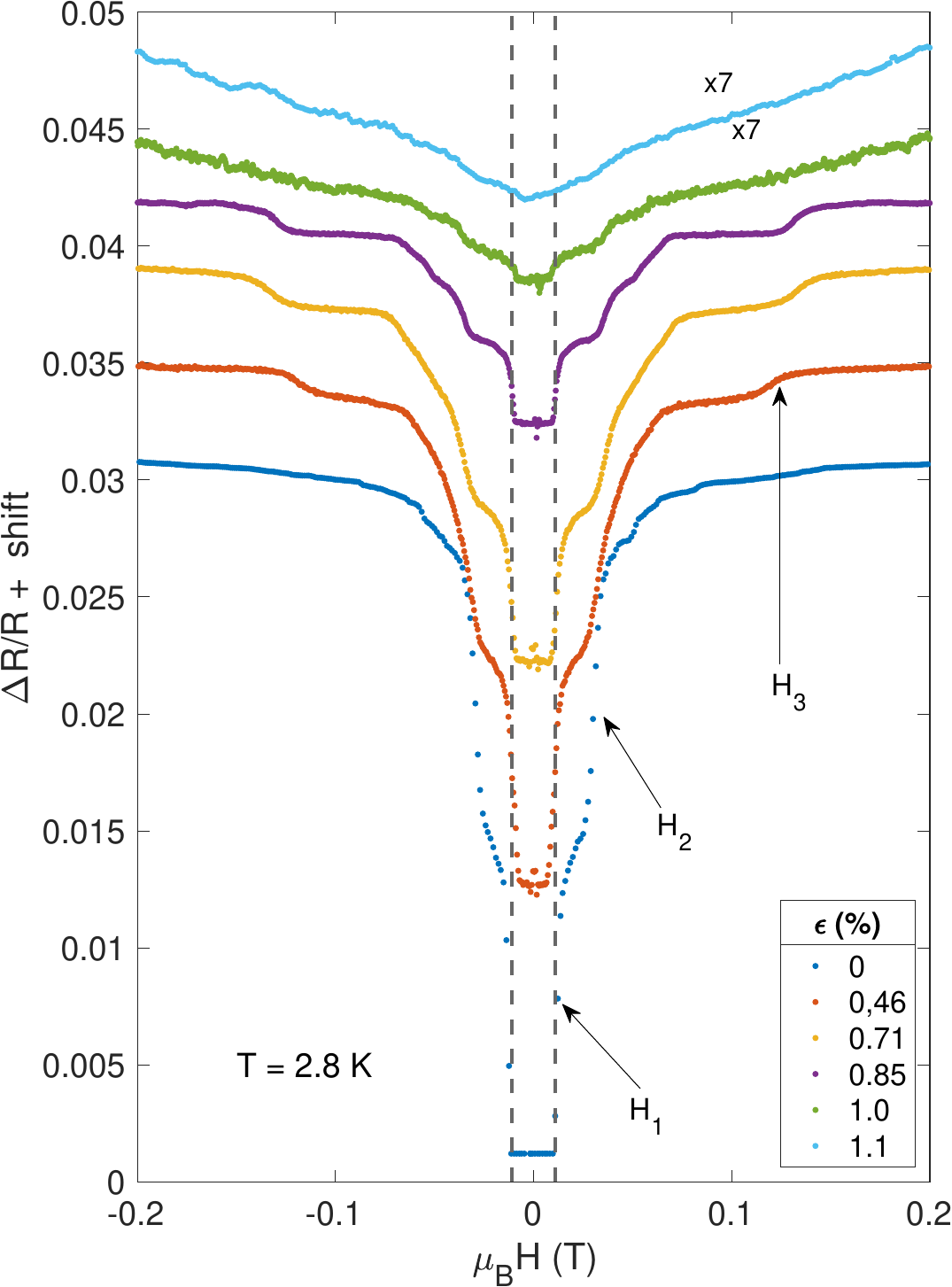}
\caption{Dependencies of resistance on magnetic field at different deformations of the sample. Vertical dashed lines denote positions of the critical magnetic field for bulk indium (see text for details).}
\label{fig:rh}
\end{center}
\end{figure}

Fig. \ref{fig:rh} shows the results of magnetoresistance $\Delta R/R\equiv [R(H)-R(0)]/R(0)$ measurements in the region of low magnetic fields at different values of strain corresponding to Fig.~\ref{fig:rt}. The dependencies are smooth but exhibit resistance steps. Position and amplitude of the steps depend on the strain value. Such smooth stepwise dependencies are observed only in devices with blurred contact boundaries which are visible in Fig.~\ref{fig:scheem}(b) as dark blue regions of the contact edges. 
With increasing strain, the relative value of magnetoresistance change decreases. In devices with sharp contact boundaries, a single sharp step-like jump to the normal state is observed.

\begin{figure}[h]
\begin{center}
\includegraphics[width=0.5\textwidth]{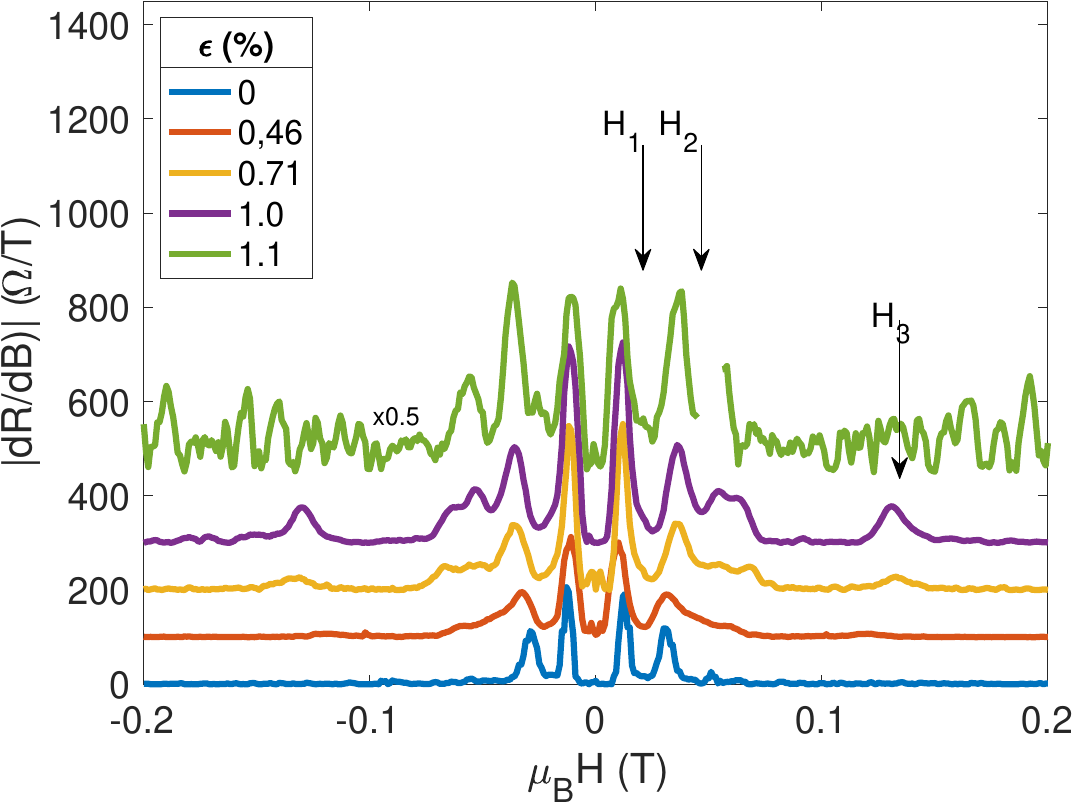}

\caption{Dependencies of the derivative modulus $|dR/dB|$ on the magnetic field at different strain levels. The colors of the curves correspond to Fig.~\ref{fig:rh}.}
\label{fig:drdh}
\end{center}
\end{figure}

In Fig.~\ref{fig:drdh},  the dependencies of the derivative modulus, $|dR/dB|$, on the magnetic field are plotted for each studied strain value. For all curves, at least two clearly distinguishable maxima of the derivatives are observed in low fields of 0.01-0.03~T,  denoted as $H_1$ and $H_2$. A maximum corresponding to $\mu_B H_3\approx 0.12$~T, where $\mu_B$ is the Bohr magneton, is also visible, which appears at $\epsilon = 0.46\%$ and disappears at $\epsilon > 0.85$\%.

\begin{figure}[h]
\begin{center}
\includegraphics[width=\linewidth]{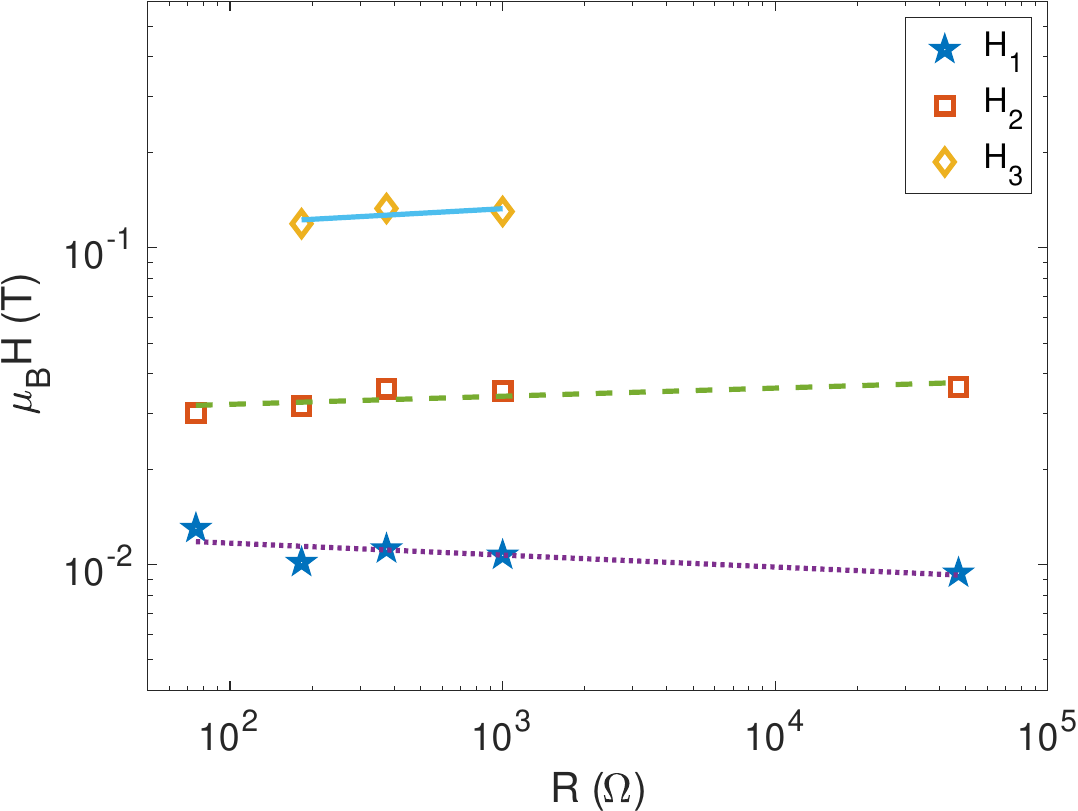}
\caption{Dependencies of the positions of the maxima $dR/dH$  on the resistance of the sample.}
\label{fig:positions}
\end{center}
\end{figure}

Fig.~\ref{fig:positions} shows the positions of the maxima of  $|dR/dB|$ as a function of the resistance of the sample. It is evident that the value of $H_1$ decreases by approximately 20\%\ with increasing strain, while for $H_2$ and $H_3$ a slight shift towards higher magnetic fields is observed.

\begin{figure}
\includegraphics[width=0.9\linewidth]{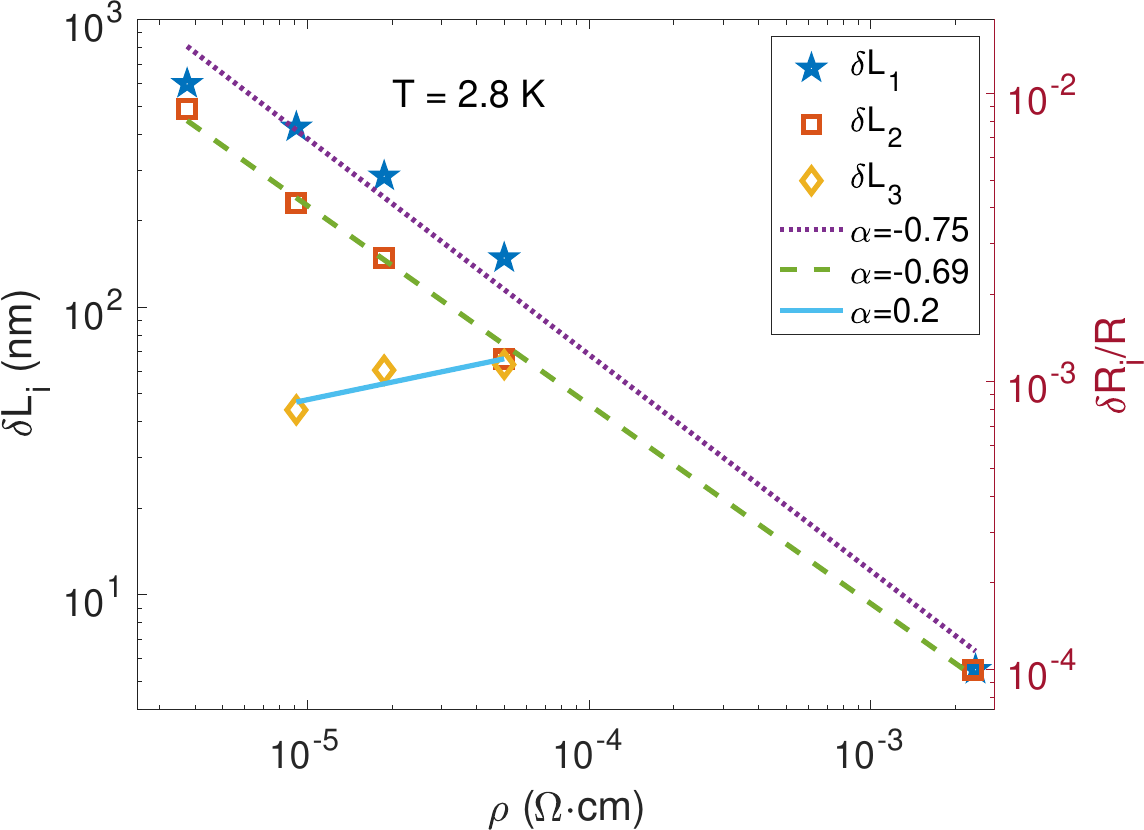}
\caption{Dependence of the relative amplitude of resistance steps (right scale) and the corresponding characteristic lengths (left scale) on the sample resistance. The straight lines show the approximation by the least squares method.}
\label{fig:lr}
\end{figure}

If we assume that the steps are associated with the disappearance of the superconducting region of the length $\delta L_i$ ($i=1,2,3$) near the edge of each indium contact, then by the magnitude of the step $\delta R_i$ we can estimate the spatial scale of the appearance/disappearance of superconductivity as $\delta L_i=(1/2)(\delta R_i/R)L$, where $L$ is the length of the sample. Fig.~\ref{fig:lr} shows $\delta R_i/R$ and the corresponding characteristic length as a function of sample resistance. In the initial undeformed state, the lengths $\delta L_1$ and $\delta L_2$ approach a micron. As the resistance of the sample increases, both $\delta L_1$ and $\delta L_2$ decrease according to a power law $\delta L_{1,2}\propto R^{\alpha}$ with $\alpha \sim -0.7$. In contrast, the value of $L_3$ depends weakly on the magnetic field and {\em increases} with sample resistance.

\section{Discussion}
The device under study is a TaSe$_3$ crystal coated with an indium film deposited on both its ends. At $T=2.8$~K, superconductivity is induced in TaSe$_3$ near the edges of the indium contacts due to the proximity effect. When a magnetic field is applied to such a device, steps in the $R(H)$ dependencies should be expected, associated with the suppression of superconductivity in different regions of the device. The purpose of the subsequent analysis is to identify the observed steps.

The first question concerns the position of the steps corresponding to the suppression of superconductivity in the indium film. To answer this question, we measured the magnetoresistance of indium films deposited using the same deposition procedure that was used to prepare TaSe$_3$ samples. It turns out that continuous films have only one sharp transition corresponding to the disappearance of superconductivity in a magnetic field $\approx 0.011$~T, corresponding to the value $H_1$. Therefore, $H_1$ corresponds to the suppression of superconductivity in the indium film.

This conclusion raises the question of the origin of the remaining steps that arise in higher magnetic fields, at which the superconductivity of the contacts is already suppressed.

As already mentioned, it turned out that to observe the described effects, contacts with blurred edges are needed, i.e. films with a boundary of smoothly decreasing thickness. The indium layer with a thickness smaller than the magnetic field penetration depth 
ensures the preservation of superconductivity in a significantly wider range of magnetic fields.  For instance, the critical magnetic field of 65 nm thickness indium film is almost ten times larger than that of bulk indium \cite{toxen1961size}.
These thinner regions remain the source of superconductivity in nearby regions of the TaSe$_3$ crystal, the suppression of which produces steps in $R(H)$.

For reliable observation of step-like magnetoresistance features at fields $H_i$ ($i=1,2,3$), a moderate amount of indium contact blurring is crucial. Minimal blurring results in an overly sharp transition, hindering the observation of superconductivity near $H_3$. Conversely, excessive blurring, leading to $\Delta R/R \gtrsim 1$ at $T=2.8$~K, obscures the step-like features due to strong $R(H)$ dependence (see Supplemental Material \cite{suppl}). An optimal blurring, yielding a few-percent $\Delta R/R$, is desired to observe a step at $H_3$. Blurring is essential for a gradual disappearance of superconductivity in a wider magnetic field range above the critical field of bulk indium, enabling step observation.

The presence of the $\delta L_i(R)$ dependencies indicates that the superconducting properties of the device are affected by the resistance of TaSe$_3$.  
It is interesting to compare the measurement results shown in Fig.~\ref{fig:lr} with the characteristic length over which the induced superconductivity extends, i.e. the coherence length $\xi$.
In the clean limit $\xi= \hbar v_{F}/2\pi k_BT$, and in the dirty limit $\xi= \sqrt{\hbar v_{F}l/6\pi k_BT}$, where $v_{F}$ is the velocity on the Fermi surface, $k_B$ is the Boltzmann constant \cite{shmidt1997_book}. In turn, $v_F\propto n^{1/d}$, where $n$ is the current carrier concentration, and $d$ is the dimension of the Fermi surface (sphere, cylinder, or plane). Assuming $n\propto 1/R$, we obtain $\xi\propto R^\alpha$, where $\alpha=-1/d$ in the clean limit and $\alpha=-1/2d$ in the dirty limit.

The slopes of the $\delta L_{1,2}(R)$ dependencies (Fig.\ref{fig:lr}) correspond to $\alpha = -0.7$, which is closest to the expected value only for the Fermi surface dimensionality intermediate between 1D ($\alpha=-1$) and 2D ($\alpha = -0.5$) in the pure limit. The fields $H_{1,2}$ correspond to the critical fields for superconductivity induced in bulk TaSe$_3$ via the proximity effect through different crystal facets.

The most interesting dependence is $\delta L_3(R)$, which is  weak and corresponds to a small {\em positive} value of $\alpha = 0.2$ (diamonds, blue straight line as a guide for eyes in  Fig.~\ref{fig:lr}). Note that the resistance steps that could be attributed to $H_3$ are absent in the initial state ($\epsilon=0$), appear at intermediate deformations ($\epsilon = 0.46\%$, $\epsilon=0.71\%$ and $\epsilon = 0.85\%$) and disappear near the metal-insulator transition ($\epsilon = 1.0$\%) (Fig.~\ref{fig:rh}). This behavior is definitely not what is expected for the bulk state.

It has been reported that TaSe$_3$ is already in the state of a strong topological insulator \cite{lin_2021-strain_TI}, or passes into this state at relatively small deformation \cite{hyun_2022_strain_TI}. 
Since the bulk states of TaSe$_3$ compete with the surface states in the proximity effect, the decrease in the density of bulk states with increasing $\epsilon$ enhances the superconductivity of the surface states.
This expectation is consistent with the behavior observed for $\delta L_3$ (diamonds in Fig.~\ref{fig:lr}). As it is clear from Fig.~\ref{fig:lr}, the absence of steps corresponding to $H_3$ at $\epsilon = 0$ does not imply the absence of the surface states, but rather may be a consequence of their small contribution, which makes them difficult to observe.

With further increase of strain, TaSe$_3$ falls into the region of a topologically trivial insulator and the surface states should disappear, which agrees with the results obtained.

It should be noted that the length $\delta L_3$ is small, which makes it difficult to carry out interference measurements similar to those used in \cite{Proxti3,Proxti4,Proxti5,hart2014induced,pribiag2015edge,Esin2023gete,kononov2020one,ZhangPRB2011_Bi2Se3sc}. This smallness may be due to the more complex structure of topologically protected states arising against the background of metallic states \cite{lin_2021-strain_TI}.

The data presented in Figs.~\ref{fig:rh} and \ref{fig:drdh} also contain other small features that may be due to the anisotropic nature of the physical properties of TaSe$_3$, but are unlikely to be reliably identified within the framework of the current study.

\section{Conclusion}
We demonstrate here the possibility of using the superconducting proximity effect as a tool for detecting surface states in topological materials using simple magnetotransport measurements of hundreds of micrometer-scale devices. 
For a response to a magnetic field to occur, a smooth decrease in the thickness of the superconductor film at contact edges must be created. Using this method, we demonstrate the presence of superconducting surface states  induced by the proximity effect in TaSe$_3$ under small stretching ($\epsilon \gtrsim 0.4$), and their disappearance when $\epsilon \gtrsim 1$\%. The results obtained confirm the conclusions of \cite{lin_2021-strain_TI,hyun_2022_strain_TI}, based on density functional calculations and ARPES measurements, about the presence
of a strong topological insulator state in TaSe$_3$ in the region of intermediate deformations before the transition of the compound to a topologically trivial insulator.

\section{Data availability} The data that support the findings of this article are openly available \cite{data}.

\bibliography{prox}

\end{document}